**Teacher Leader Identity Development in a Physics Teaching Community of Practice: A Narrative Case Study**

Hamideh Talafian*, Maggie Mahmood, Eric Kuo, Morten Lundsgaard, and Tim Stelzer

Affiliation for all authors:

University of Illinois Urbana-Champaign

Department of Physics

1110 W Green St

Urbana, IL, 61801

*Corresponding Author's Email: talafian@illinois.edu

**Funding**

This research was funded by NSF DRK-12 grant number 2010188.

**Acknowledgments**

The authors would like to thank the teachers of the professional development program for participating in this research.

## Abstract

This longitudinal narrative case study examines the leadership identity development of Kayla, a female high school physics teacher, across five years of participation in a community of practice (CoP). Drawing on multiple data sources, we trace how her participation evolved from classroom-level practice to broader institutional influence. Four interconnected dimensions of leadership identity development were identified: personal, social, practice, and institutional culture change. Findings show how CoP membership provided collective legitimacy and relational support that enabled Kayla to move beyond identity negotiation toward transformation — successfully advocating for an AP Physics course and increasing female student participation. This study contributes to teacher professional learning literature by demonstrating the value of an identity lens for understanding teacher leadership as a gradual, socially enabled process. The longitudinal design reveals transformative conditions that cross-sectional studies often obscure, highlighting narrative case study as a productive methodology for capturing teacher leadership identity development over time.

***Keywords:*** teacher leadership identity; community of practice; longitudinal narrative case study; professional development; physics education; high school

## Introduction

The impact of science teacher professional development (PD) is often evaluated through their instructional practices in the classroom (e.g., Thompson et al., 2016; Watkins et al., 2017). While this is one window into teachers' developing pedagogical knowledge, instructional goals, self-efficacy, and agency, it ignores other potential aspects of teacher professional development. One alternative approach focuses on science teacher leadership development, how teachers take up professional roles in their school and beyond (Howe & Stubbs, 2003; Taylor et al., 2011; York-Barr & Duke, 2004) following PD encouragement and support.

In this study, we have adopted a more holistic view of teacher leadership development, examining it through the lens of identity development, which encompasses both how they see themselves and how they are perceived by others. Rather than equating leadership with assuming formal roles outside the classroom (Wenner & Campbell, 2018), we draw on Dozier's (2007) definition of STEM teacher leaders as those who influence others. This influence manifests in teachers' classroom practices and extends beyond the classroom, encompassing a wide range of activities that can shape the work of fellow educators. Shifts in instructional practice, for instance, can themselves be understood as a form of leadership, reflecting the impact of professional development within the classroom and rippling into broader school and professional communities.

Grounded in this perspective, we examine Kayla's teacher leadership development, a high school physics teacher, through the lens of identity development within a physics teaching Community of Practice (CoP) (Wenger, 1998). We conceptualize leadership development as a process of identity formation situated within a disciplinary context and unfolding across both school-based and broader professional communities. Prior research has demonstrated the role of CoPs in shaping teachers' identities through meaning-making and the exchange of strategies (Weinberg et al., 2021). Teachers who participate in these communities bring diverse backgrounds and, in doing so, co-construct communities that often transform their professional identities. By framing leadership development as an identity trajectory, rather than a set of enacted roles, we aim to offer a more comprehensive account of how teachers change through PD and how those changes reverberate through the communities they serve.

Drawing on multiple data sources and narrative inquiry methods, we constructed a longitudinal profile of Kayla's leadership identity development. This approach led us to the following research question: *How does Kayla narrate and enact the development of her leadership identity within the physics teaching CoP and her school context?*

Findings from this study suggest that leadership development in disciplinary spaces, such as a physics teaching CoP, cannot be disentangled from physics teacher identity development. Studying leadership in such contexts, therefore, requires sustained attention to the intersection of disciplinary identity construction and leadership trajectories

## Theoretical Underpinning

To develop the conceptual framework for this work, we drew on multiple theoretical approaches to create a comprehensive map for studying leadership identity development within a community of practice. This framework (see Figure 1) is used for guiding our data analysis.

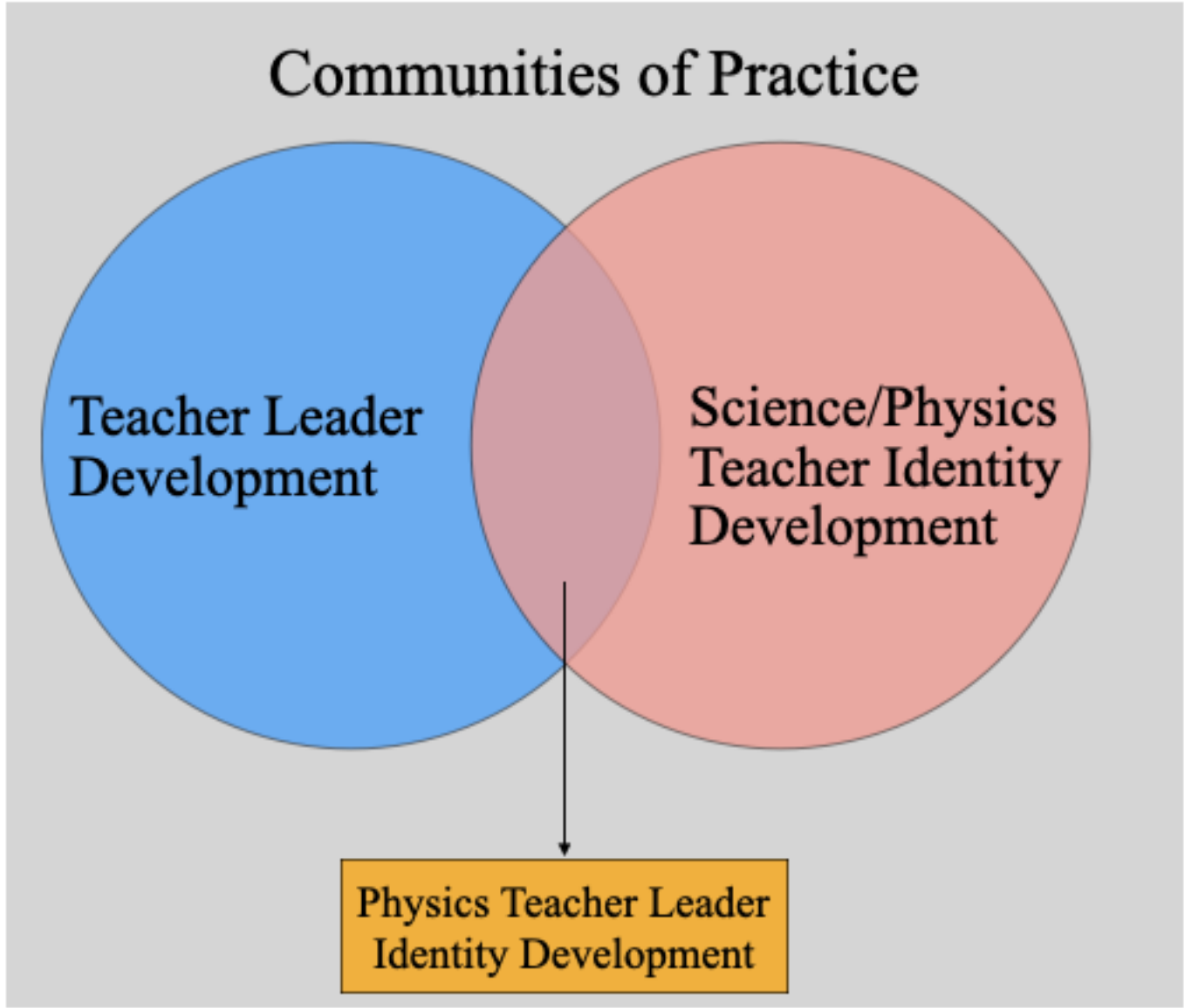

***Figure 1.*** Conceptual Framework

### Physics Teacher Leader Identity Development

Physics teacher leader identity development lies at the intersection of teacher leader development and science/physics teacher identity development (See Figure 1; our Conceptual Framework). Rather than treating leadership as separate from disciplinary teaching, our framework conceptualizes physics teacher leader identity as emerging through the integration of leadership practices and disciplinary identity development in a CoP.

***Teacher Leadership Development***

Teacher leader development frameworks vary in how narrowly or broadly they conceptualize leadership development. In this study, we intentionally draw on frameworks that align with our broader definition of teacher leadership, which extends beyond formal positions. Specifically, the Dempsey metaphors of leadership (Dempsey, 1992) and the Kentucky Teacher Leadership Framework (Criswell, 2016; Kentucky Teacher Leadership Work Team, 2015) emphasize leadership enacted through instructional practice, collaboration with colleagues, and engagement at the school and community levels, rather than leadership defined solely by designated roles or titles. These frameworks foreground teachers' influence on teaching and learning through everyday practice and participation. This approach is well-suited for examining physics teacher leadership when the goal is not merely to take formal leadership roles but to impact others through classroom practice.

In Dempsey's four metaphors of leadership development, teacher leadership has been described as a trajectory of an ever-expanding sphere of influence, starting inside the classroom (fully functioning person and reflective practitioner) and moving outside the classroom (scholar and learning partner) (Dempsey, 1992). Similarly, the Kentucky Teacher Leadership Framework (Kentucky Teacher Leadership Work Team, 2015) outlines six levels of leadership development that begin in the classroom. The first level, *leading from the classroom*, emphasizes instructional decision-making grounded in best practices in teaching and responsive to students' needs. Leadership then expands outward through increasing levels of influence: collaborating with peers, leading small groups, participating in school- or district-level decision-making, engaging in regional or national leadership networks, and ultimately fostering broader change through partnerships with the community, business, and industry, as well as fundraising efforts (Criswell, 2016).

### ***Science/Physics Teacher Identity Development***

In the preface to *Studying Science Teacher Identity: Theoretical, Methodological, and Empirical Explorations*, Avraamides draws on Wenger to argue that examining teacher change through the lens of identity development constitutes an ontological approach to learning –one that foregrounds how learning reshapes who teachers are, not merely what they know or do (Avraamidou, 2020; Wenger, 1998). From this perspective, identity development captures changes in teachers' ways of being as they participate in professional and disciplinary practices and bring their prior experiences to the science classroom (Avraamidou, 2014, 2020; Gee, 2000; Luehmann, 2007; Moore, 2008).

Across the literature, teacher identity studies, despite conceptual and theoretical variations, converge on the understanding that teacher identity is socially constructed, dynamic, multilayered, and multifaceted, often encompassing multiple, intersecting sub-identities (Avraamidou, 2020). Within science education specifically, scholarship has extended this work by attending to reformed-minded teaching orientations (Avraamidou, 2014) and the roles of emotion and power within scientific and educational communities (Avraamidou, 2016; Zembylas, 2005). Yet the field has not fully explored discipline-specific dimensions of teacher identity, as it does for student science/physics identity development (Carlone & Johnson, 2007;

Hazari et al., 2010). Some of the explored constructs capture how science teachers position themselves in relation to scientific knowledge, authority, and purpose—for example, how they perceive the status and power of their work and how these perceptions shape their instructional practices (Luehmann, 2016). To fully capture teacher leader identity development, it is essential to attend to both leadership and science identity aspects.

**Leadership Development: A Process of Identity Change in a Community of Practice (CoP)**

Identity development, being recognized by oneself or by others as a particular "kind of person" (Gee, 2005; Luehmann, 2007), does not occur in isolation. Within the Communities of Practice (CoP) framework (Lave & Wenger, 1991), identity is understood as both a personal and social transformation shaped through participation in a community. Individuals make meaning by negotiating their roles, values, and practices, through which they simultaneously influence and are influenced by others in the community (Palus & Drath, 1995; Wenger, 1998). Identity development is thus a process of *becoming*, involving moments of identity negotiation, resistance, and claiming that emerge through interaction with others. For teachers, this process is reflected in the stories they construct about their professional lives (Clandinin et al., 1999) and in the communities in which they participate, learn, and grow (Wenger, 1998); together, these experiences shape who they understand themselves to be.

From this perspective, studying teacher leadership requires an identity-based lens that attends to the communities that nurture teachers' development, both within and beyond formal program structures. An identity lens situated within CoP enables a richer understanding of how leadership practices develop through participation, recognition, and evolving relationships. It also highlights how communities shape teachers' identities, and how teachers, in turn, influence others and the community itself. Examining leadership development in this way is therefore essential to understanding the program's impact on teachers' leadership trajectories.

## Methods

### Study Context

The context of this study is a partnership program between a large R1 university and a group of high school physics teachers in a Midwestern state. The central goal of the program is to increase access to high-quality physics instruction by supporting teachers as they adapt their practice to changes in Advanced Placement (AP) physics exams and align their curricula with the Next Generation Science Standards (NGSS). To achieve these goals, the program integrates curricular resources with sustained professional development (PD), designed to respond to teachers' diverse contexts and needs (Talafian, Lundsgaard, Mahmood, Shafer, et al., 2025).

The curricular component provides teachers with free access to university-level physics materials, including an online platform and a laboratory device currently used in the university's introductory physics courses. The PD component connects teachers with a broad network of peers and university faculty through a blended model of virtual and in-person sessions, totaling approximately 100 hours per year per teacher for up to four years. Ongoing online PD supports teachers in adapting and implementing these materials in their classrooms. Each year, a new cohort joins returning teachers, enabling cross-pollination of curricula, activities, and pedagogical approaches. The focal participant of this study, Kayla, joined the program in its second year.

Over seven years, the program has grown from four to 65 teachers across 60 schools, with PD structures evolving in response to participants' curricular needs and feedback. Teachers in the program have between 0 and 31 years of teaching experience, and approximately half teach outside their original disciplinary field. Participating schools are similarly diverse, with enrollments ranging from 135 to 4,200 students; roughly half are designated Title I schools. Rural, suburban, and urban schools are all well represented.

**Study Design**

We employed a narrative inquiry (Clandinin & Caine, 2008) case study design to develop Kayla's case. Since leadership identity is central to this study and identity research often emphasizes individuals' stories of lived experience, we adopted narrative inquiry (Riessman, 2008) for this case study to foreground Kayla's stories. Narrative inquiry attends to how individuals interpret their experiences and how those interpretations shape who they become (Riessman, 2008). As Kerby (1991) notes, "we invent who we are in the stories that we tell" (p. 6). Rather than treating stories as fixed or purely descriptive accounts, narrative inquiry understands them as *co-constructed* between participants and researchers and as continually shaped by context, relationships, and time (Clandinin et al., 1999; Riessman, 2008). Narrative inquiry emphasizes particular events, contexts, and meanings, while also acknowledging that one person's story can illuminate broader social patterns (Chase, 2003; Kim, 2015). In this way, Kayla's story not only reveals her own leadership identity development but also reflects experiences that may resonate with other teachers whose voices are less often captured.

**Case Selection**

Kayla is a science teacher with 12 years of experience, including 7 years teaching multiple levels of physics, including Advanced Placement (AP) Physics. She is now in her third year of teaching AP Physics at a large public high school with about 2,000 students, 51% of them classified as low income, and 63% of whom are from minoritized backgrounds. Kayla's academic background is in chemistry, with a minor in Secondary Education, and she has earned physics endorsements that qualifies her to teach both general and AP-level Physics.

Kayla was selected for our study based on the significant changes observed during her participation in the program. In May 2019, she was asked to teach physics after the previous physics teacher retired. Although she was initially reluctant and declined the offer, the administration could not find another qualified teacher in the district, so Kayla ultimately took on the role. Kayla subsequently joined our Physics Teaching Community of Practice (CoP) in 2021, through which she built skills and confidence that prepared her to teach AP Physics. Despite experiencing impostor feelings, managing personal health challenges, and navigating the pandemic, she remained committed to improving her practice. In 2023, she received approval to teach AP Physics while simultaneously completing her master's degree in physics education. Currently, she is the only physics teacher at her school and teaches multiple levels of physics. Over time, Kayla went from a reluctant teacher to one who initiated major reforms in her school's physics program and attributed her success to her participation in the program. Figure 1 shows the timeline of key events in Kayla's career developed from the interviews.

**Figure 1**

*Timeline of Key Events for Kayla's Case.*

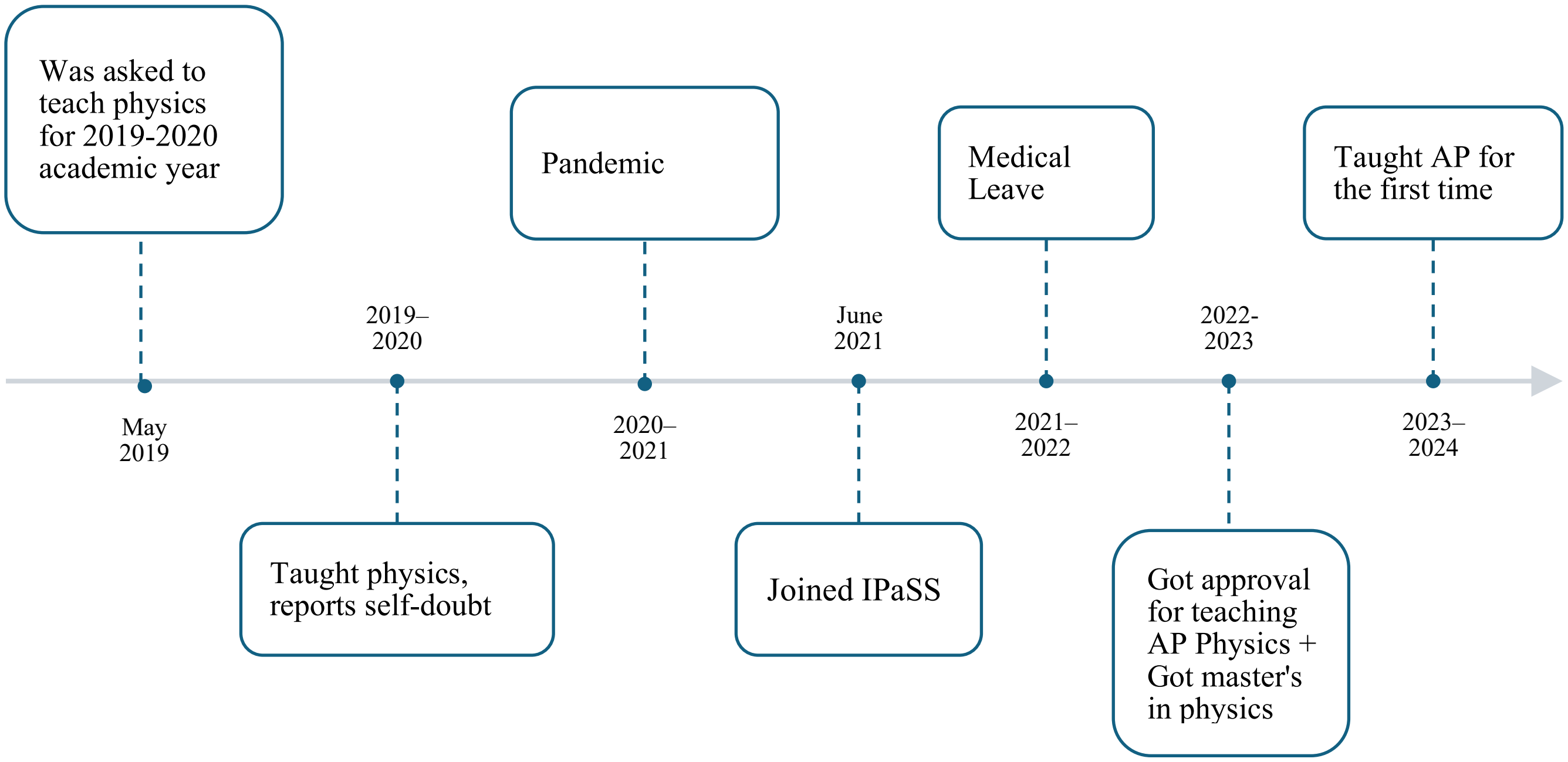

**Data Sources**

For the creation of Kayla's case, both qualitative and quantitative data were collected over the six years of her participation in a program from 2021 to 2026. Quantitative measures were embedded to support interpretation and triangulation, not to generate independent claims. Table 2 shows the year of collection, type, and details of the data collected from Kayla to create this longitudinal case study.

**Table 2.** *Data Sources for Kayla's Narrative Development*

| Type of Data | Data Collected | Year of collection | Details |
|---|---|---|---|
| Interview with Kayla | Interview 1 | 2022 | CoP Involvement and experiences |
| | Interview 2 | 2022 | CoP goals |
| | Interview 3 | 2024 | Leadership Identity Development (LID) |
| | Interview 4 | 2025 | Momentum Unit |
| | Interview 5 | 2026 | Member Checking |
| Interview with program facilitator | Facilitator Interview | 2024 | Others' view of Kayla's LID |
| Interview with Kayla's colleagues in CoP | CoP teacher Interview 1 | 2024 | Interview with Francesca |
| | CoP teacher Interview 2 | 2024 | Interview with Amy |
| Survey | Baseline Demographic Survey | 2021 | Demographics and teaching experiences |
| | Teachers' Self-Efficacy Beliefs (TEBS) Survey | 2023 | A validated survey given in a retrospective manner |
| | Perceived Organizational Support (POS) | 2023 | A validated survey was given twice, once about their school support and once about program support |
| | Guided Inquiry Survey (GIS) | 2023 | A validated survey given in a retrospective manner to measure lab instruction philosophy changes |
| Other Data | Personal Value Narrative | 2023 | A reflective matrix that allows people to document their professional goals |
| | Summer Development Plan | 2021-2025 | A reflective form asking teachers what they intend to change in their teaching and how. |
| | Emails to program facilitators | 2022-2024 | Updates about her teaching (e.g., getting approval to teach AP) |

***Interviews.*** The primary sources of qualitative data were multiple interviews with Kayla, an interview with one program facilitator (author 2), and interviews with two additional teachers in the program. These interviews provided insight into how Kayla perceives herself, how the program facilitator perceives her growth, and how her colleagues perceive her development—three perspectives that are cited as crucial for studying identity development (Gee, 2005; Wenger, 1998) from a multi-faceted perspective. However, narrative inquiry interview methods differ in important ways from other qualitative, interview-based approaches.

In narrative inquiry, interviews are treated as extended conversations rather than structured question-and-answer exchanges (Riessman, 2008). Participants are encouraged to lead with their stories, take longer turns, and move naturally from one story to another, allowing the researcher to "follow participants down *their* trails" [italic in the original text](Riessman, 2008,

p. 24). Although this conversational style requires giving up some of the efficiency associated with structured interviewing, it yields richer, more meaningful data (Devereux, 1967). While careful transcription and coding remain important, narrative inquiry cautions against treating transcripts as the sole source of evidence. Rissman (2008) emphasizes that narrative researchers draw on observations, ongoing relationships, and conversations that unfold over time. For this reason, although Table 2 summarizes the data used in constructing the narrative, Kayla's case is grounded in five years of sustained engagement with her.

***Surveys.*** Survey data included a baseline demographic survey and three validated surveys, all administered to Kayla in summer 2023. The surveys measured self-efficacy, perceived organizational support, and teachers' lab philosophy.

The Teachers' Self-Efficacy Beliefs (TEBS) survey (Dellinger et al., 2008) is a tool that measures teachers' self-efficacy beliefs across six constructs: communication and clarification, classroom management, accommodating individual differences, student motivation, managing learning routines, and higher-order thinking. The items specifically ask teachers whether they can create a situation that reflects these constructs. Examples of communication are: "I can communicate to students content knowledge that is accurate and logical." Or "I can clarify students' misunderstandings or difficulties in learning." An example of accommodating individual differences is: "I can implement teaching methods at an appropriate pace to accommodate differences among my students." This survey was administered to all participating teachers, and Kayla's results were extracted for this study. We administered the survey retrospectively, meaning teachers needed to reflect on each item and rate it for "before joining the program" and "after joining the program."

The Perceived Organizational Support (POS) instrument (Eisenberger et al., 1986) was adapted and administered twice in June and August 2023. In June, the items were modified to assess participants' perceptions of support from their own organizations/schools, and in August, the same items were adapted to assess support from the program. This allowed us to quantitatively compare the two sources of social support. Sample items were: "My school values my contribution to its well-being", "My school fails to appreciate any extra effort from me."

Guided Inquiry Survey (GIS) (Cheung, 2011) was another validated survey that was given to Kayla. This survey was initially used to trace chemistry teachers' perceptions of changes in how they implement labs. We modified this survey for physics teachers and intended to see whether their philosophy of teaching labs changes over time toward a more open-ended approach. The goal was to capture changes in teachers' practice. Sample items were asking "Designing experiments should be a formal part of physics curriculum", Guided inquiry experiment is better than structured inquiry experiment to get students to develop a wider range of practical skills."

All three validated surveys were administered retrospectively (Moore & Tananis, 2009; Pratt et al., 2000), meaning that the teachers needed to reflect on the survey questions both prior to and after joining the program.

***Other Qualitative Data Sources.*** In developing Kayla's narrative inquiry case, three other qualitative sources of data helped: 1) Summer development plan, 2) Personal value narrative, and 3) Emails to program organizers. The summer development plan was simply a form given to teachers every year to record which changes they wanted to make to their course, how they wanted to make those changes, and a reasonable timeline for completing the prep needed to make those changes possible. In Kayla's case, we reviewed four documents from 2021 to 2025. She did not attend the in-person workshops in 2025; hence, there is no updated summer development plan for her this year.

The personal value narrative (Wenger et al., 2011) is a reflective tool designed to assess value creation within communities. It helps teachers consider how their participation has influenced them professionally and socially, how they have adapted their practices, and the impact they have had on their environment. These aspects align closely with our focus on leadership identity development from personal, professional, social, and practical perspectives. See Appendix A for Kayla's personal value narrative.

Emails to program organizers were also another e source of qualitative data. Kayla sent those impromptu emails to us either in appreciation or in recognition of some of the program's support. These emails helped both in identifying the case and in learning about the direct ways Kayla is being impacted by the program.

## Findings

Using a broader definition of leadership development and incorporating an identity development lens, this section presents findings of Kayla's trajectory of leadership identity development. Findings are not only showing Kayla's self-perceptions of herself as a physics teacher but also examine how others perceive her, offering a comprehensive view of her trajectory of change over the course of her participation in the program.

### Trajectories of Physics Teacher Leadership Identity Development

In response to the research question of "How does Kayla narrate and enact the development of her leadership identity within the physics teaching CoP and her school context?" we list four dimensions of narration and enactment that map out changes: 1) personal, 2) social, 3) practice, and 4) culture change. The dimensions of leadership identity development in Kayla's case are analytically distinct but empirically interconnected. As a result, some patterns of change appear across dimensions, introduced in one section and developed more fully in others.

***1) Personal Dimension: From "Imposter" Feelings to Legitimate Self-Positioning***

Longitudinal analysis of survey, interview, and observational data—both from Kayla and her colleagues —revealed a steady shift in confidence in how Kayla perceives herself and in how she is perceived as a legitimate physics teacher leader. When Kayla entered the program, she described herself as feeling like a "fraud," a perception that persisted throughout her first year. This impostor positioning manifested in limited interaction with peers, silence during group discussions, and a tendency toward isolated work.

> … at first kind of almost feeling I was a fraud in the program, which is all in my head. um and so just over time, seeing the others where they struggle on some times with the content or sometimes with getting students to engage with materials, things like that, and I think there's just a commonality that we've started to find over time and, especially being in the smaller group. (Interview 1)

These feelings of inadequacy, however, did not originate with the program. They can be traced back to Kayla's entry into teaching physics in 2019, when she was asked to take over the course following the previous teacher's retirement. She initially expressed a clear preference for someone else to teach physics. When the administration was unable to find a replacement, she agreed to teach the course, but described her first year as "struggling with self-doubt" and uncertainty about pacing. At that time, her goal was simply to "get through this class."

> And so the 2019- 2020 school year was a bit rough because I was teaching this course for the first time. And my pacing was very off. because I thought my students would pick up on things faster. And so there's a lot of struggling with self-doubt, with not really being

> sure of pacing it correctly. And at the same time there were certain topics that I was having to teach myself, because my physics experience in college was not the best. (Interview 3)

After two years of teaching physics, Kayla joined the program in June 2021, marking a turning point in her understanding of herself as a physics teacher.

> So I read the email [invitation from CoP colleague]. And I was like, this is interesting. and so like, I decided, why not? I'll, you know, apply for it. And I, honestly, I was like, I'm not going to get accepted. Obviously, that did not come true. But basically what happened as a result of that is… it allowed me to start building a network with other physics teachers. (Interview 3)

Through sustained participation, she reported that she "actually started really liking teaching physics" and noted, "I think just over time this program is allowing me to gain some confidence in my own ideas." This growing confidence made her more inclined to share her work and to recognize her own strengths in teaching physics and the "valuable things" she could contribute to the community.

> I feel like over the past 3 years I've gotten a lot more confidence in my own teaching and so I'm a lot more willing to share about my experiences. Because there was that fear at first where it's like, um, and it sometimes creeps in where I'm like, am I doing anything right with my students? And so… But I've had a lot more confidence to be able to talk about my experiences with my physics classes and saying, hey, we did this. It really worked! They really enjoyed it! …. But it's also the case that what didn't work for their students may work for mine, and vice versa. And so, just being a little bit more willing to share some of the stuff that I've done and seeing this is what worked for my students

> [that] may not work for yours, or you can adjust it however you want. But this is something that I found valuable that I did. And so, just remembering that I do have strengths in my teaching and I do have valuable things to share with the group. (Interview 3)

Similarly, Kayla's Teachers' Efficacy Beliefs System (TEBS; Dellinger et al., 2008) results in 2023 show notable growth from before joining the program to after, particularly across constructs tied to instructional decision-making and student sensemaking (Table 3). Using a four-point scale (1 = very weak beliefs; 4 = very strong beliefs), Kayla demonstrated gains in self-efficacy in communication and clarification practices, accommodating individual differences, and fostering higher-order thinking skills. Yet, scores related to student motivation and managing learning routines remained stable, suggesting that changes were aligned with program emphases.

**Table 3.** *TEBS Survey Results for Kayla.*

| | Before joining IPaSS (mean score) | After joining IPaSS (mean score) |
|---|---|---|
| *Communication/Clarification* (aka monitoring & feedback for learning) (9 items) | 3 | 3.8 |
| *Management/climate* (aka classroom management and maintaining a positive classroom climate) (10items) | 3.1 | 3.4 |
| *Accommodating individual differences* (aka planning & accommodating for individual differences) (7 items) | 2.7 | 3.4 |
| *Motivation of students* (3 items) | 3.6 | 3.6 |
| *Managing learning routines* (3 items) | 3.3 | 3.3 |
| *Higher order thinking skills* (4 items) | 2.2 | 3.2 |

The quantitative TEBS results are aligned with our qualitative interview data, showing Kayla's confidence reflected in making principled instructional decisions rather than merely implementing prescribed curricula. Notable examples include changes to the sequence of teaching units and the reversal of the order of concepts in the momentum unit. “At some point in there I started doing impulse change of momentum first because I felt like that was more logical.” Here, she notes that students now understand “why we can have change in momentum and why we can also have conservation of momentum,” and can connect changes in momentum to external forces. These observations show that Kayla is questioning the conventional sequencing as she gets more confident. We will cover more details about this aspect under the *Practice Dimension: Enacting Leadership Through Pedagogical Agency* section.

The shift in Kayla’s confidence was also evident to her colleagues. Two teacher colleagues in the program similarly recalled her as initially quiet and reserved. Francesca, who had entered the program a year before Kayla and had invited her to join, described noticeable changes in Kayla’s participation over time. She recalled that Kayla initially appeared “quiet but soon started finding her space in the program” and became “naturally more vocal and open,” particularly when leading sessions. Francesca’s observations highlight a transition from deference to a confident teacher, reflecting external recognition of Kayla’s growing sense of legitimacy alongside her evolving role in the community. Although uncertainty is still being narrated, Kayla reframes it as a pedagogical resource:

> I started feeling like my own past experiences, my own feelings with inadequacy towards it [teaching] really helped my students because I wasn’t up there thinking, why aren’t you getting this? It’s like, it’s okay. Physics is hard. But we’re going to try doing this problem this way instead. We’re going to try doing something else. (Interview 3)

To conclude, Kayla’s increased levels of confidence, reflected in pedagogical agency, position her as a legitimate physics teacher and are a clear departure from impostor narratives. In the following section, we examine how Kayla's emerging identity gets leveraged through participation in the physics teaching CoP.

### 2. *Social Dimension: From Peripheral Participation to Influential Membership in CoP*

The data show that joining the CoP was a turning point in developing Kayla's leadership identity. Every time Kayla talks about challenges and opportunities, she mentions the CoP as a place that enabled positive change. Below, we examine two prevalent narratives that highlight the social aspect of leadership identity development within the CoP for Kayla.

**2.1 Shifts in Participation Patterns: From Support-Seeking to Contribution.** Early in the program, Kayla perceives the CoP as an opportunity to "start building a network with other physics teachers" and to gain "instant feedback" about instructional challenges. At this stage, her participation centered on observing peers' practices mostly in silence, learning new problem-solving techniques, and drawing on curricular resources that colleagues "so willingly" shared. Hence, the CoP was a resource for learning rather than a site of contribution. Yet, her participation narrative shows that this pattern changes in silence by observing the shared struggles of her peers:

> There was something so different about being able to get this instant feedback, and to hear what other teachers were doing, what they were struggling with. And it helped a lot to hear that. This other teacher in a school district that has completely different socioeconomic status or demographic than mine, they're struggling with the same things. And so it's like obviously, I have to work to make things work for my kids. But it's really neat to hear how they're trying to make things work for their kids. And so that was a huge change for me. To be able to have those conversations. And even just sometimes where I just listen in, I'm like, that's a really good idea, that's interesting. And so like, I'll

> reflect on, okay, can I implement this in some way in my classroom? How would that look like for me? (Interview 3)

Over time, Kayla's participation shifted toward more reciprocal engagement. Within the CoP, Kayla started adapting others' materials as opposed to using materials as they are (e.g., "electric house project"). At this stage, she also contributed her own instructional resources:

> So, my students were kind of struggling with the idea of practical acceleration, and one of my students was like, I want to do a lab with paper airplanes. and he was like, "Can you make a lab with paper?" And I developed a lab. It's not a perfect lab, by any means, but it gets them the idea of acceleration. Especially the idea of accelerating being a concept where it's changing the last few years slowing down. And so I created that accelerating airplane lab which I'm pretty sure is in the [CoP program] folder. (Interview 3)

Even though Kayla still finds herself "socially awkward" in the CoP, she finds the in-person meetings great for professional dialogue: "there's something vastly different about being able to communicate face-to-face." Additionally, she finds online meetings as a place where "We bounce ideas off of each other." Both of which illustrate active engagement in collective meaning-making in the community. As a contributing member, she now volunteers to share pedagogical ideas or experiences in online meetings and leads some of the meetings. These patterns indicate movement along a trajectory of participation, from peripheral engagement toward more central involvement in the community.

**2.2 Contributions Recognized by Peers in the CoP.** As Kayla's participation became more reciprocal, she was increasingly recognized by peers as a legitimate and valued contributor to the community. Recognition of Kayla's contribution was triangulated with facilitator observations and colleague interviews. For instance, Francesca, in her interview, explicitly acknowledged Kayla's contributions: "Kayla has taken on leadership roles that we like to see within our professional development in terms of running the sessions and being, you know, a leader, especially in terms of things related to equity pieces." This statement directly reflects Francesca's perception that Kayla is fulfilling a legitimate leadership function within the CoP.

Further, Francesca emphasized the community's interest in learning from Kayla's work, particularly her efforts to develop an AP Physics program and increase girls' enrollment in physics. She noted that "other people appreciate that Kayla is able to be very concrete about what she's trying—this is what's worked and this is what she's tried that hasn't worked in her own space." Francesca further emphasizes that Kayla's transparency on "sharing struggles" has been even more helpful than watching polished presentations that lead to feelings of inadequacy in teaching. This emphasis on transparency and specificity framed Kayla as a credible source of practical insights, a form of peer legitimation in the CoP.

Similarly, Amy's interview underscores Kayla's recognition as a legitimate teacher leader, drawing on both her interactions within the CoP and her perspective as a parent whose son was in Kayla's class. Amy highlighted Kayla's contributions to the CoP, such as running meetings and volunteering to lead workshops. Similar to Francesca, Amy particularly appreciated Kayla's practical insights into classroom practices and navigating challenging situations at school. She also emphasized Kayla's effectiveness in advocating for instructional and structural change at her school, especially in securing approval for AP Physics 1. These statements show how Kayla's colleagues considered her a legitimate physics teacher leader once her participation patterns changed. Amy's parent perspective is particularly important here, as she observed Kayla's influence extending beyond her own classroom when university-based materials were adopted in the freshman physics curriculum, reaching even students Kayla did not directly teach.

Together, these peer perspectives illustrate how Kayla's leadership was recognized not only through participation but through her capacity to influence decisions and effect change, further legitimating her leadership identity within and beyond the community.

### 3. *Practice Dimension: Enacting Leadership Through Pedagogical Agency*

Kayla's leadership identity became more visible in action when all sources of data directed us to Kayla's changes of teaching practices. Interview data with Kayla indicated several high-level changes in both instruction and assessment. This data showed that her teaching has moved toward a more student-centered approach, informed by physics-specific pedagogies introduced during the professional development sessions, particularly the Investigative Science Learning Environment (ISLE)[1] framework. She also reported changes in assessment practices, including the use of project-based assessments and allowing retakes, reflecting a shift toward more flexible and learning-oriented grading practices.

**Figure 2**

*Changes in Kayla's Leadership Identity in Practice Dimension.*

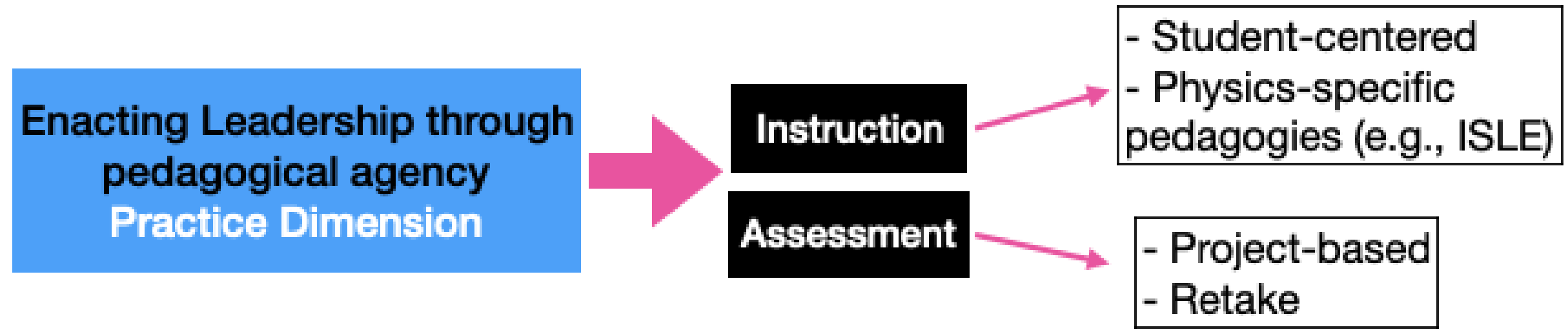

**3.1 Instructional Changes.** Kayla's narrative in all interviews showed some degree of change in instruction and assessment. In particular, in interview 4 about the momentum unit,  she gave more details about how she has decided to change different aspects of her instruction in this unit. Initially, Kayla described her first attempts at teaching momentum with a degree of uncertainty and reliance on established resources (e.g., Pasco lab). She mentions that her initial unit plan for momentum was based on what “others had done that I had seen online.” Here, she describes early uncertainties:

> So, looking back at it [first year of teaching physics, before joining CoP], because I know I did like *nothing* that’s actually on this sheet [curriculum]. But the thought at the time was to start off with a lab investigation in which they were going to see when two objects collide together, how the initial velocity and mass of one affect the velocity of the two objects after the collision. So, the original idea was because there’s this argument driven inquiry lab, which is a whole like set of books to talk about, then they were going to have to actually come up with their own investigation to do it. I know that I did not end up doing this partially, because I realized by then that my students did not have the capability of doing this well. And so it ended up being that we did a lab that Pasco had, where we looked at how things ended up hitting and colliding, and then doing some analysis with the collision. (Interview 4)

[1] **Investigative Science Learning Environments (ISLE)** (Etkina & Van Heuvelen, 2007) is a student-centered instructional approach that mirrors the processes of scientific inquiry, where learners actively construct and test models through experimentation, evidence-based reasoning, and iterative refinement rather than passively receiving information.

Her narrative suggests a lack of a fully developed personal pedagogical approach. In the context of the guided-inquiry lab, she seems to find it challenging to facilitate the process, noting that students may lack the necessary foundational skills to succeed in that format. She also admits later in the same interview that “I know, I need to hit these things [learning goals], but the logical progression was not necessarily known to me, because I never had taught it” further highlights her early uncertainty regarding the optimal sequencing and pedagogical flow of the unit.

Later in her narrative, she articulates some of her observations of students’ performance, which led to redesigning the momentum unit. She observes that students “put numbers into an equation" but "don't necessarily actually know how to problem solve" or "get the actual concepts behind it." Kayla explains a pedagogical decision that allowed changing the sequence of lessons in a unit—*micro-sequencing*. So, instead of teaching the momentum concept first, she chose to introduce impulse first "because I felt like that was more logical." This decision reflects a move away from compliance-based teaching toward internally justified pedagogical reasoning. Importantly, Kayla attributes this shift to observing how other program fellows restructured their units in CoP. This further highlights the role of CoP in her decision-making process that we reviewed previously.

Another example of instructional change is in Kayla's first year of teaching AP Physics, when she deliberately changed the sequence of the momentum unit to follow the order of work, energy, and power—*macro-sequencing*. She justified this macro-sequencing decision by noting that students needed prior conceptual grounding in energy to understand elastic collisions in the momentum unit (See Figure 3). Such decisions signal growing confidence (as discussed under Personal Dimension) not only in content knowledge, but as a legitimate physics teacher with the authority to shape curriculum in the service of student understanding.

**Figure 3**

Macro-sequencing Shift in Kayla's Physics Units.

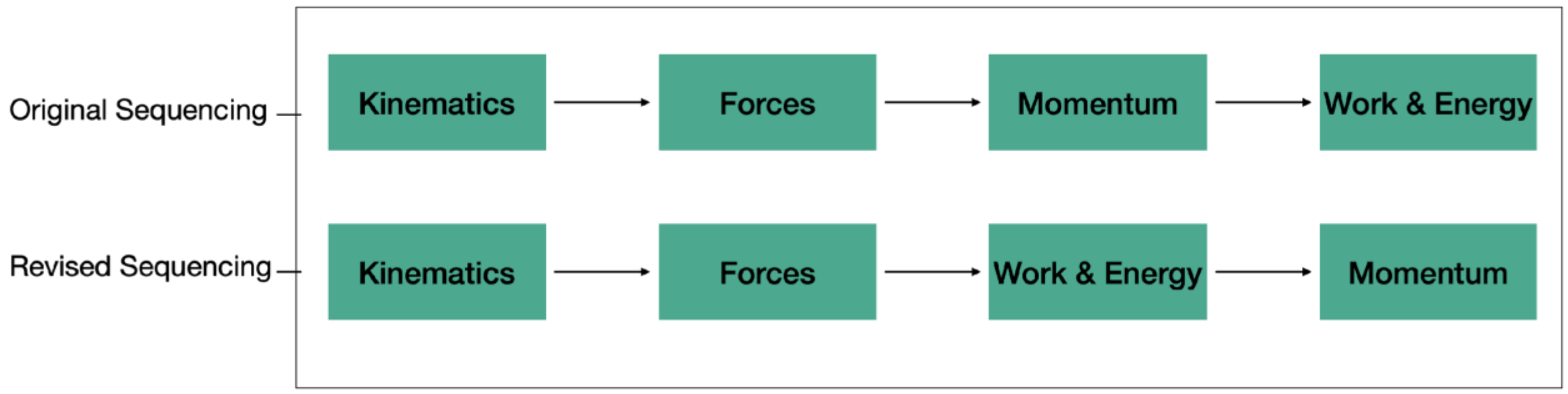

Later in interview 4 (momentum), she reflects on another example of a big shift in her instruction– *systems thinking*. Kayla explicitly mentions that "the systems thing has actually been huge" for her teaching of energy. Systems thinking in physics is an approach to understanding physical phenomena by focusing on interactions, relationships, and dynamics of the whole system rather than its individual components or equations. Kayla mentions that this change came from her interactions with the CoP. "I think emphasizing those different systems has been really important to my students because they seem to think like everything that we've learned for physics has to be true in every single circumstance…whereas before [joining CoP] I feel like it was just like physics is true, but there are sometimes where it's not." This quote shows that the CoP has impacted her thinking about physics, and through which she has impacted her students. Hence, Kayla mentioned systems thinking as one of the most valuable changes in her instruction.

All three examples of changing instruction—micro- and macro-sequencing and systems thinking —illustrate shifts in Kayla's developing leadership identity in the CoP, as reflected in her daily practices.

**3.2 Assessment Changes.** Another aspect of change in the practice domain of Kayla's leadership identity development is assessment changes. Kayla's narrative shows a shift toward project-based assessment and more flexible grading or "retakes." Both of these assessment changes have been adopted from the CoP teachers in the program. In regard to project-based summative assessments, she names two specific projects that she's adopted from two veteran teachers in the community: 1) the electricity house, 2) the paper roller coaster project. These two projects are both designed and implemented by veteran teachers and were examples of non-traditional assessments. These projects were introduced in the PD meetings, and Kayla decided to adopt them for her classes. In her narrative, there was an extensive explanation of the adoption and implementation process, where she was sharing her computer screen with the interviewer and was walking us through the details. She emphasized the importance of making connections between school physics and real-world applications for the students. She also mentioned how she turned part of the project into an alternative assessment to avoid offering a multiple-choice test.

Regarding retakes, Kayla relies on Joe Feldman's book *Grading for Equity* (Feldman, 2023) in implementing remedial assessment gradually. In line with Feldman's approach, retakes were not treated as a second chance for points, but as an opportunity for students to demonstrate their eventual mastery of content**,** emphasizing what students know at the end rather than when they learned it. Observing increased student success, Kayla gradually expanded this practice to other courses she taught.

> I was one of the first people in the school to start doing assessment retakes and having students do remediation, so going back and looking at what they got wrong and figuring out why they got the question wrong. And so I started slowly, like, incorporating it into more classes, because I first did it in the 2019-2020 school year. So, I slowly started incorporating it into my classes that I was then… teaching where other teachers were also teaching it, because I'd implemented it in physics first, because I was the only one teaching that course. And so… What we started seeing is we started seeing more student success, and then… Eventually, the school [adopted it] which wasn't just because of me, it was also because they had gone through… similar processes of reading. I think it's Joe Feldman's book, grading for equity, we started trying to do it, but a lot of the teachers initially were, you know, grumbling about it, and I was like, well, I've got almost all of mine built into my class. Like, it's just natural that's what happens, because I want my students to be able to understand something more at the end. And so, that's what some of the other teachers have started doing, too, because they're like, well, then that makes

> sense, because then they're forcing their students to do it, and it makes sense for the classes where the content really builds. (Interview 5)

Kayla's narrative on retake shows that she eventually influenced the school to adopt a similar method. We will review this aspect under the Culture Change domain in more detail, especially as it is linked to an environment where the colleagues were hesitant to implement the retakes.

All in all, our data suggest that Kayla has been able to shift toward a more agentic physics teacher, making more informed instructional decisions. While she acknowledged that it wasn't an "entire pedagogical shift", because she was already experimenting with some changes, the CoP facilitated the implementation of her evolving pedagogical philosophy: "the program allowed me the space and the tools to finally put those in practice in my class."

### *Culture Change Dimension: Exercising Leadership Within Institutional Constraints*

Kayla's leadership identity is particularly visible in her narrative, which has resulted in shifting the school culture around female participation, equitable assessment, and positive perceptions of physics, despite persistent institutional and interpersonal constraints. Across interviews and colleague accounts, her actions reflect strategic, sustained leadership. Three interrelated sub-dimensions characterize this domain.

**4.1 Navigating Institutional Constraints Through Strategic Persistence.** Kayla's narrative shows significant institutional barriers in expanding and reforming the physics program. Yet the data from interviews with her and her colleagues show that she has been able to overcome these challenges through strategic persistence, backed by the support of the CoP. The most notable of these challenges is the multi-year process of getting approval for the Advanced Placement (AP) Physics courses in her school. Beginning in 2019 (prior to joining the program), repeated requests were met with administrative delay, indicating that “we’ll just wait.” Progress occurred only after a shift in leadership, coupled with Kayla’s involvement in the CoP, which “pushed it forward.”

In the process of getting approval, resource cost and time constraints further complicated the process. In response to district resource costs, Kayla strategically leveraged a free, College Board–approved textbook and mentioned the CoP’s support to the school board for implementation and instructional support. Kayla’s colleague in the CoP, Amy, also recognized this leverage and attributed it to Kayla’s persistence and analytical approach in the face of challenges. “She is very analytical. She is very well-spoken, and she will come up with airtight arguments backed up by data [and] student responses… to the point where you just can’t argue with her.” As to the time constraints, even though Kayla was given 40 hours of paid curriculum preparation time in the summer, she still felt a need for more support. To overcome this institutional constraint, she strengthened her participation in the CoP, which enabled collaborative planning and support from colleagues in aligning her practices with College Board standards.

Eventually, Kayla got approval for teaching AP Physics at her school in 2023 and wrote to us saying: “After I tried for 3.5 years, last night, the school board approved AP Physics 1 as a new course starting for the 2023-2024 school year. I’m especially grateful for the CoP [the program’s name] because it makes this transition feel far less intimidating!”

Kayla’s narrative of navigating the challenges of securing approval for AP Physics courses illustrates how she exercised leadership within institutional constraints. While support from her school was present, the CoP played a more significant role in enabling these changes. Results from the Perceived Organizational Support (POS) survey reflect this difference, showing high perceived support from the CoP program (M = 4.25/5) compared to more moderate support at the school level (M = 3.45/5). Although school support was limited, it was not absent; Kayla was granted curricular autonomy, flexibility in assessment practices (e.g., retakes as opportunities for demonstrating improved understanding), and some dedicated preparation time (i.e., 40 hours). Together, these findings suggest that Kayla learned to navigate institutional constraints by combining persistence, strategic framing, and leveraging external support systems such as her participation in CoP.

**4.2 Negotiating Collegial Resistance and Divergent Instructional Norms.** In addition to structural constraints, Kayla navigated resistance and skepticism from colleagues whose priorities and pedagogical norms differed from her own. A recurring tension centered on the school's emphasis on supporting the "bottom 20%" of students, often at the expense of advanced learners. Kayla framed this imbalance as a systemic equity issue, arguing that high-achieving students were being underserved and advocating for AP Physics as a corrective.

Colleagues also sometimes minimized the complexity of her teaching context, characterizing her classes as easier because they consisted primarily of the same level students, "honors, AP, or elective students." Kayla countered this narrative by emphasizing the distinct challenges faced by high-achieving students, including stress and mental health concerns.

As discussed in the previous section, Kayla described taking instructional risks by introducing inquiry-based, hands-on activities (e.g., the water balloon drop lab) alongside assessment retakes and remediation practices, even while facing ongoing institutional and collegial constraints. She noted that some of these practices had previously been abandoned by other teachers due to negative experiences, making their reintroduction feel like a challenge to existing norms around rigor, grading, and classroom management. Reflecting on this, Kayla shared:

> I've had so much criticism from some of my coworkers being like, 'Why are you doing the test retakes, blah blah blah?' And I know that they work with a different group of students than the ones that are taking the elective for physics, but at the same time it's like, why would you not want to offer up that chance, when you get something like that, where that applies for more than just physics. (Interview 2)

Despite this pushback, Kayla continued to implement these approaches in her own classroom, drawing on what she observed in her students' learning and engagement. Over time, she saw shifts not only in her classroom but also more broadly, as practices such as retakes began to gain acceptance and were eventually adopted at the school level. These examples illustrate how Kayla exercised leadership by modeling alternative practices, grounding change in evidence of student learning, and remaining steady in the face of initial resistance, qualities that Amy also identified in her interview when describing Kayla.

**4.3 Cultivating a Positive Culture Around Physics Learning.** In Kayla's narrative, there are multiple examples of creating a positive culture around learning physics. One example was creating the growth mindset culture in relation to retakes in her school:

> I mean it's such a social emotional learning piece to teach them that you know you may not immediately get something. And you have the option built in that it's okay to not immediately get something and that your teachers going to work with you, and you know if you fail the test it's not the end of the world because your teachers contact you and or come up to you in class and be like all right, you didn't get it the first time, so what do we need to do? And having that conversation with them and so kind of building that culture

> out of having it where they can develop a growth mindset, without even mentioning those two words together by having that culture, set up in the first place. So, like that's kind of been my goal, and what's happened as a result, as I think there's a lot of people that are like how I was in college, that all of a sudden say, you know what? I'm going to take this class, because I do want to go into engineering, and I should take physics now because [Ms. X] is going to make sure that I have every opportunity to succeed, so basically the word is getting down to the younger students. And some of them, they had me as freshmen on and so that helps because you know they're seeing that culture set up now. But the word does get around where they're like you know the class is challenging but whether it's going to make sure that you can succeed in it. Um. So that's kind of been where I've come from with that. (Interview 2)

Another example of creating a positive culture around physics learning was Kayla's effort to dismantle gender and racial barriers by creating a more inclusive culture. She inherited a program characterized by traditional, exclusionary practices that had contributed to the avoidance of physics among girls and students from historically minoritized backgrounds. "I've been trying to implement some of the STEP-UP[2] program practices from APS to get more females interested in physics, as the 2018-2019 physics classes were 95% male." In response, Kayla made a deliberate effort to "build a culture of positivity" around physics through adding more hands-on projects as summative assessments, and allowing for revisions of assessments, i.e., retakes; both of which she believed addressed female students' test anxiety and increased their participation and performance. Kayla's reform efforts and explicit messaging that "physics *is* for everybody" led to significant shifts in enrollment, retention, and student attitudes. Her motivation for reform was consistently framed in equity-oriented terms both in her narrative and in her colleagues'. Although direct empirical data from her school are not available to substantiate these outcomes, indirect evidence from Kayla and her colleagues' interviews suggests that these changes were meaningful and likely associated with her instructional approach.

[2] The STEP-UP program, developed by the American Physical Society, utilizes research-based curricula to help students—particularly girls and those from historically minoritized backgrounds—develop a "physics identity." Key practices include the "Women in Physics" and "Careers in Physics" modules, which are designed to counter cultural stereotypes by highlighting diverse representation and communal career values, thereby shifting the classroom from a gatekeeping environment to one of belonging.

The culture change dimension of Kayla's leadership development reveals that institutional constraints do not stop her but motivate her to support students. Her trajectory of leadership identity development showed that Kayla identified gaps in opportunity, articulated a responsibility to challenge systemic patterns, and derived legitimacy from student outcomes, particularly students' increased confidence and preparedness for AP exams. This growing internal sense of legitimacy was reinforced by external recognition from the CoP and allowed her to work against institutional constraints to protect students' interests. Her email to the program organizers captures how her professional identity got leveraged with the CoP support in an institution where she is the only physics teacher:

> My students are learning. They're getting used to learning concepts through multiple methods. I'm trying something new in a couple weeks with a lab practicum where they get to try to drop a water balloon on a math teacher's head. Even those that aren't sure why they took this course are getting into it. I'm excited about this course, even if it's a big adjustment for me personally. The only reason why this has been a success so far is because of the investment from the [university CoP] partnership. It genuinely has made me a better teacher, and I have been able to learn so many new strategies to take back to my own classroom. The PLC like nature is also invaluable as a singleton teacher. I'm just feeling very grateful this Friday, and I wanted to pass it along. -Kayla

### *Analytic Considerations for Findings*

An alternative narrative of Kayla's trajectory may have omitted her identity development and her evolving participation in the teacher CoP. A focus on Kayla's evolving pedagogical practices could attribute her teacher development to (I) her growing confidence as her physics teaching experience increases and/or (ii) her growing content and pedagogical knowledge, to which her participation in a teacher PD program certainly contributed. However, we aimed to show that a focus on Kayla's leadership identity development captures meaningful dimensions of Kayla's professional growth. Explicit consideration of Kayla's leadership in improving educational practices and opportunities at her school captures her growing self-efficacy and locus of agency. An identity lens also brings focus to the identity development trajectory that can be seen through Kayla's participation in the physics teacher CoP, through how Kayla positions herself and how she is perceived by others. Through this narrative, we aim to show the potential value of focusing on teacher leadership identity development as a method for capturing typically unmeasured dimensions of teacher growth and professional development.

## Discussion

In this narrative inquiry study, we looked at Kayla's leadership development through the lens of identity change in the CoP. Using a broader definition of leadership development and multiple data sources, we adopted a multifaceted view of leadership development that encompasses professional identity change within the CoP. To build Kayla's case, we looked at her involvement with the physics teacher CoP over five years and how Kayla has perceived herself and how others have perceived her development.

Taken together, we illustrated changes through four interrelated dimensions: 1) the personal dimension which showed changes in Kayla's confidence, 2) the social dimension which reviewed changes as a result of and in connection with the CoP, 3) practice dimension, reviewed changes in teaching practices both in instruction and assessment, and 4) changes in school culture which included creating a positive culture around learning physics, recruiting more female students, and creating AP Physics course to support students' participation. Below, we examine the main findings in relation to the broader body of literature on teacher leadership identity development.

### *Communities of Practice Enabling Teacher Leadership Identity Development*

This narrative case study shows that Kayla's leadership identity development was supported by her participation in a physics teacher Community of Practice (CoP) at a critical moment in her career. Early in her trajectory, Kayla faced professional isolation, disciplinary uncertainty, and limited informal support, often seeking help through online resources (e.g., Facebook). The CoP provided timely access to a physics-specific professional community, offering both social and intellectual support that aligned with her immediate needs. In-person participation was particularly significant, as it created relational entry points and a sense of belonging that supported her engagement and confidence over time. This result aligns with prior findings on the connections between socio-emotional peer support and teacher disciplinary engagement (Mahmood et al., 2024) and instructional change within the CoP (Talafian, Lundsgaard, Mahmood, Kuo, et al., 2025).

Importantly, the CoP did more than improve Kayla's instructional practices; it enabled her leadership by providing legitimacy through colleague recognition, which resulted in increased confidence and elevated her self-perception of legitimacy. While Kayla's school offered some structural support, such as curriculum autonomy, preparation time, and flexibility around assessment, these supports became actionable largely through the CoP. Some CoP program structures, including pairing her with senior colleagues (e.g., Francesca), inviting her to co-present, and positioning her as a facilitator in meetings, created increased recognition from the community.

These findings align with prior research emphasizing the role of social participation in shaping teacher identity and practice (Avraamidou, 2014; Carlone & Johnson, 2007; Deneroff, 2016; Gee, 2000; Zhai et al., 2024). For example, Deneroff (2016) argues that teacher practice is not primarily driven by knowledge or beliefs, but rather by socially constructed identities developed through participation in CoP. From this perspective, professional development becomes impactful when it supports identity transformation rather than simply transmitting knowledge. They further argue that identity provides agency, enabling teachers to resist and reshape institutional norms (Deneroff, 2016). This framing is similar to what we found about Kayla's resistance in the face of institutional constraints, as we see Kayla's leadership identity development through CoP participation as she works to change her school culture and norms.

### *From Negotiation of Identity to Transformation*

Kayla's narrative is an example of transformation of leadership identity development rather than a negotiation of one's identity in the face of challenges. Previous studies, such as Upadhyay, have talked about the negotiation of identities. Both Upadhyay (2009) and this study examine how teachers navigate institutional and cultural constraints, but they differ in what that navigation ultimately leads to. In Upadhyay's work, the teacher (Maria) exercises agency by adapting her instruction and negotiating tensions between her beliefs and dominant norms. However, this work remains largely confined to Maria's classroom, and identity development reflects ongoing negotiation rather than broader change.

In contrast, this study shows how such negotiation can evolve into transformation when supported by community and recognition. Kayla's participation in CoP provided legitimacy and collective support, enabling her to extend her influence beyond her classroom. Practices that initially faced resistance became more widely accepted, pointing to shifts not only in her teaching but in her school context. This comparison highlights a key contribution of this study: it identifies the conditions that enable identity to move beyond negotiation toward transformation. These transformative conditions, rooted in the relational resources of the CoP, suggest that when teachers have access to sustained collegial support and spaces of legitimate participation, leadership identity development can move from a defensive negotiation of constraints toward an expansive reshaping of institutional possibilities.

**Limitations and Conclusions**

This study highlights the complex nature of a psychological construct, leadership identity development, which is inherently intertwined with other overlapping constructs, e.g., self-efficacy (Talafian, 2020). Our longitudinal narrative case of Kayla demonstrates how these constructs unfold over time in personal, social, practice, and cultural dimensions. While our analysis focused on these dimensions, we acknowledge that other important constructs, such as competence, frequently emphasized in science identity frameworks, were implicitly tracked. Instances of sequencing in the practice dimension speak to the development of competence in Kayla's trajectory.

The depth and nuance captured in this study were made possible through carefully designed research methods. We believe narrative inquiry was particularly suited to exploring the multifaceted evolution of Kayla's leadership identity, allowing us to trace not only her development but also potential changes in her colleagues' perspectives and behaviors. Nevertheless, we recognize that this approach may not be applicable to all research contexts.

Hence, we encourage future research to use the results of this study to design tools for capturing leadership identity changes and measuring them in a larger population.

Finally, our work underscores the value of longitudinal, in-depth, case-based approaches for understanding how leadership emerges under constraints and within complex institutional contexts. Comparative case studies could further elucidate how leadership identity trajectories differ across STEM disciplines, providing richer insight into the interplay between individual growth, practice, and PD designs in teaching CoP contexts.

## References

Avraamidou, L. (2014). Studying science teacher identity: Current insights and future research directions. *Studies in Science Education*, *50*(2), 145-179.

Avraamidou, L. (2016). *Studying science teacher identity: Theoretical, methodological, and empirical explorations.* (L. Avraamidou, Ed.). Sense Publishers.

Avraamidou, L. (2020). Science identity as a landscape of becoming: Rethinking recognition and emotions through an intersectionality lens. *Cultural Studies of Science Education*, *15*(2), 323-345.

Carlone, H. B., & Johnson, A. (2007). Understanding the science experiences of successful women of color: Science identity as an analytic lens. *Journal of Research in Science Teaching: The Official Journal of the National Association for Research in Science Teaching*, *44*(8), 1187-1218. https://doi.org/https://doi.org/10.1002/tea.20237

Chase, S. E. (2003). Taking narrative seriously: Consequences for method and theory in interview studies. *Turning points in qualitative research: Tying knots in a handkerchief*, *3*, 273-298.

Cheung, D. (2011). Teacher beliefs about implementing guided-inquiry laboratory experiments for secondary school chemistry. *Journal of Chemical Education*, *88*(11), 1462-1468.

Clandinin, D., & Caine, V. (2008). Narrative Inquiry. In L. M. Given (Ed.), *The Sage Encyclopedia of Qualitative Research Methods* (Vol. 1, pp. 542-545). SAGE Publications, Inc. https://doi.org/http://dx.doi.org/10.4135/9781412963909.n275

Clandinin, D. J., Connelly, F. M., & Bradley, J. G. (1999). Shaping a professional identity: Stories of educational practice. *McGill Journal of Education*, *34*(2), 189.

Criswell, B. (2016). *Kentucky Teacher Leadership Framework*. Collaborative for Teaching and Learning (CTL). Retrieved January 27 from https://ctlonline.org/kentucky-teacher-leadership-framework/?utm_source=chatgpt.com

Dellinger, A. B., Bobbett, J. J., Olivier, D. F., & Ellett, C. D. (2008). Measuring teachers' self-efficacy beliefs: Development and use of the TEBS-Self. *Teaching and Teacher Education*, *24*(3), 751-766.

Dempsey, R. (1992). Teachers as leaders: Towards a conceptual framework. *Teaching Education*, *5*(1), 113-122.

Deneroff, V. (2016). Professional development in person: Identity and the construction of teaching within a high school science department. *Cultural Studies of Science Education*, *11*(2), 213-233.

Devereux, G. (1967). *From anxiety to method in the behavioral sciences*. De Gruyter Mouton.

Dozier, T. K. (2007). Turning good teachers into great Leaders. *Educational Leadership*, *65*(1), 54-59.

Eisenberger, R., Huntington, R., Hutchison, S., & Sowa, D. (1986). Perceived organizational support. *Journal of Applied psychology*, *71*(3), 500.

Feldman, J. (2023). *Grading for equity: What it is, why it matters, and how it can transform schools and classrooms*. Corwin Press.

Gee, J. P. (2000). Chapter 3: Identity as an analytic lens for research in education. *Review of research in education*, *25*(1), 99-125.

Gee, J. P. (2005). *An introduction to discourse analysis: Theory and method*. routledge.

Hazari, Z., Sonnert, G., Sadler, P. M., & Shanahan, M. C. (2010). Connecting high school physics experiences, outcome expectations, physics identity, and physics career choice: A gender study. *Journal of Research in Science Teaching*, *47*(8), 978-1003.

Howe, A. C., & Stubbs, H. S. (2003). From science teacher to teacher leader: Leadership development as meaning making in a community of practice. *Science Education*, *87*(2), 281-297.

Kentucky Teacher Leadership Work Team. (2015). *Kentucky teacher leadership framework.* https://eric.ed.gov/?id=ED572280

Kerby, A. P. (1991). *Narrative and the self*. Indiana University Press.

Kim, J.-H. (2015). *Understanding narrative inquiry: The crafting and analysis of stories as research*. Sage publications.

Lave, J., & Wenger, E. (1991). Learning in doing: Social, cognitive, and computational perspectives. *Situated learning: Legitimate peripheral participation*, *10*, 109-155.

Luehmann, A. L. (2007). Identity development as a lens to science teacher preparation. *Science Education*, *91*(5), 822-839. https://doi.org/https://doi.org/10.1002/sce.20209

Luehmann, A. (2016). Practice-linked identity development in science teacher education: GET REAL! Science as a figured world. In *Studying science teacher identity: Theoretical, methodological and empirical explorations* (pp. 15-47). SensePublishers Rotterdam.

Mahmood, M. S., Talafian, H., Shafer, D., Kuo, E., Lundsgaard, M., & Stelzer, T. (2024). Navigating socio-emotional risk through comfort-building in physics teacher professional development: A case study. *Journal of Research in Science Teaching*, *61*(9). https://doi.org/https://doi.org/10.1002/tea.21949

Moore, D., & Tananis, C. A. (2009). Measuring change in a short-term educational program using a retrospective pretest design. *American Journal of Evaluation*, *30*(2), 189-202.

Moore, F. M. (2008). Positional identity and science teacher professional development. *Journal of Research in Science Teaching: The Official Journal of the National Association for Research in Science Teaching*, *45*(6), 684-710.

Palus, C. J., & Drath, W. H. (1995). *Evolving leaders: A model for promoting leadership development in programs*. Center for Creative Leadership.

Pratt, C. C., McGuigan, W. M., & Katzev, A. R. (2000). Measuring program outcomes: Using retrospective pretest methodology. *American Journal of Evaluation*, *21*(3), 341-349.

Riessman, C. K. (2008). *Narrative methods for the human sciences*. Sage.

Talafian, H. (2020). *Tracking Perceived STEM Career Identity in Minoritized Middle Schoolers in an Informal Learning Environment: A Mixed-Methods Study*. Drexel University.

Talafian, H., Lundsgaard, M., Mahmood, M. S., Kuo, E., & Stelzer, T. J. (2025). Teachers' experiences with taking an open-ended approach in teaching labs in high school physics classes. *Physical Review Physics Education Research*, *21*(1), 010140.

Talafian, H., Lundsgaard, M., Mahmood, M. S., Shafer, D., Stelzer, T., & Kuo, E. (2025). Responsive Professional Development: A Facilitation Approach for Teachers' Development in a Physics Teaching Community of Practice. *Teaching and Teacher Education*, *153*(104812). https://doi.org/https://doi.org/10.1016/j.tate.2024.104812

Taylor, M., Yates, A., Meyer, L. H., & Kinsella, P. (2011). Teacher professional leadership in support of teacher professional development. *Teaching and Teacher Education*, *27*(1), 85-94.

Upadhyay, B. (2009). Negotiating identity and science teaching in a high-stakes testing environment: An elementary teacher's perceptions. *Cultural Studies of Science Education*, *4*(3), 569-586.

Weinberg, A. E., Balgopal, M. M., & Sample McMeeking, L. B. (2021). Professional growth and identity development of STEM teacher educators in a community of practice. *International journal of science and mathematics education*, *19*(Suppl 1), 99-120.

Wenger, E. (1998). Communities of practice: Learning as a social system. *Systems thinker*, *9*(5), 2-3.

Wenger, E., Trayner, B., & De Laat, M. (2011). Promoting and assessing value creation in communities and networks: A conceptual framework.

Wenner, J. A., & Campbell, T. (2018). Thick and thin: Variations in teacher leader identity. *International Journal of Teacher Leadership*. https://scholarworks.boisestate.edu/cifs_facpubs/202/

York-Barr, J., & Duke, K. (2004). What do we know about teacher leadership? Findings from two decades of scholarship. *Review of educational research*, *74*(3), 255-316. https://doi.org/https://doi.org/10.3102/00346543074003255

Zembylas, M. (2005). Discursive practices, genealogies, and emotional rules: A poststructuralist view on emotion and identity in teaching. *Teaching and Teacher Education*, *21*(8), 935-948.

Zhai, Y., Tripp, J., & Liu, X. (2024). Science teacher identity research: A scoping literature review. *International journal of STEM education*, *11*(1), 20.